\documentclass[12pt]{book}

\usepackage[dvips]{graphicx,color}
\usepackage{makeidx,tsukuba}

\makeauthorindex
\makeindex

\newcommand{\gr}{$\gamma$-ray \,}
\newcommand{\grs}{$\gamma$-rays \,}

\begin{document}

\BookTitle{\itshape The 28th International Cosmic Ray Conference}
\CopyRight{\copyright 2003 by Universal Academy Press, Inc.}
\pagenumbering{arabic}

\chapter{TeV Gamma-Ray Observations and the Origin of Cosmic Rays III.}

\author{H.J. V\"olk\\
{Max-Planck-Institut f\"ur Kernphysik, D-69029 Heidelberg, Germany}
}

\section*{Abstract} 

\noindent The present final part of a triad of plenary talks on the results of
high energy gamma-ray astronomy and the origin of Cosmic Rays is primarily
devoted to the physics interpretation. I will start with the stereoscopic method
of operating several imaging atmospheric Cherenkov telescopes in coincidence as
pioneered by HEGRA, and a summary description of the results obtained. Then I
will turn to the search for gamma rays from Supernova Remnants over the last
decade and argue that only the quantitative comparison of observations with a
consistent theory of particle acceleration can lead to a generally acceptable
picture of Cosmic Ray origin. The elements of this theory are outlined. It is
subsequently used to model SN~1006 and Cassiopeia A, the two sources that until
now could be investigated to the required detail. The analysis shows for the
first time that emission spectra and morphological detail are in agreement with
the concept that these two distinctly different objects are representative
members of a suspected Supernova source population in the Galaxy. The continuing
study of this population is an essential part of the program of the major new
gamma-ray instruments.


\section{The HEGRA Stereoscopic System}

\noindent Representing also the HEGRA experiment in this overview I want
to emphasize first the significance of the stereoscopic method in
ground-based \gr astronomy. This observation technique positions several
telescopes in the Cherenkov light pool on the ground so that the same
gamma-ray shower in the atmosphere is observed from different positions, in
a manner analogous to the practice of a land surveyor (Fig. 1). The
reconstruction of the \gr direction is achieved on a purely geometrical
basis by combining the different shower images into a single focal plane
detector ("camera"). The shower impact point on the ground is reconstructed
similarly from the images in the spatially seperated telescopes. An
immediate result is increased angular resolution and a higher energy
resolution. At the same time it is possible to strongly improve background
rejection, in particular against local muons from unrecognized showers in
the distant atmosphere. The energy threshold for such a system is therefore
lower than for each of the individual telescopes seperately.

\begin{figure}[htbp]

\centering

\vspace{-1.5cm}

\includegraphics[width=\linewidth]{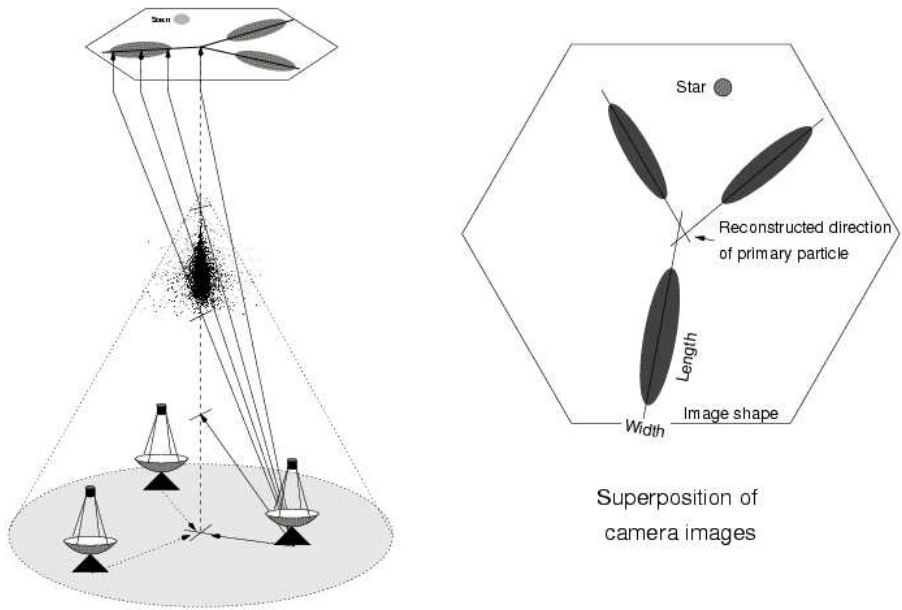}
\caption{Schematic of stereoscopic \gr observations. The Cherenkov light of the
shower electrons illuminates a cone whose basis on the ground has a radius of
about 120 m. Pointing telescopes inside the Cherenkov cone see the shower under
different angles (left panel). The elongated shower images in the focal plane (with
a length exceeding their width) point towards the primary \gr direction. Bringing
the various images into a single camera allows a geometric reconstruction of the
\gr direction. An analogous image extrapolation from the seperated telescopes
yields the shower impact point on the ground (right panel). Courtesy G.
P\"uhlhofer. \label{figure1}}
\end{figure}

The stereoscopic array of the HEGRA experiment on La Palma which was finally
dismantled in late 2002 has been combining five relatively small telescopes
of 3.3 m effective diameter ($8.5 {\mathrm m}^2$~mirror area) each. Using the
technique described above it was able to detect and thereby to confirm all the
Northern Hemisphere sources originally detected by the (single) 10 m ($75
{\mathrm m}^2$~mirror area) Whipple telescope in Arizona. In addition, the
HEGRA system detected a number of very weak sources like the first
unidentified TeV source TeV J2032+4130 in the Galaxy, the first Northern
Hemisphere Supernova Remnant Cassiopeia~A (Cas~A; Fig.2), and the first radio
galaxy in the TeV range: M87 in the center of the Virgo cluster. Consistent
with their fluxes of merely several percent of the Crab Nebula, the three new
sources could not yet be confirmed by other instruments.

I shall not enter into a discussion about the overall observational results
of TeV \gr astronomy which the previous speakers, T.C. Weekes and T.
Kifune, have already presented so ably for the Extragalactic and the
Galactic sources.  My assignment here is rather to interprete these results
with regard to the origin of Cosmic Rays (CRs). More specifically I shall
ask the question, whether and to which extent the TeV-detection of
Supernova Remnants (SNRs) in our Galaxy has solved this problem which has
eluded physicists for more than ninety years, since the discovery of Victor
Hess in 1912.

\begin{figure}[htbp]

\centering

\includegraphics[width=0.9\linewidth]{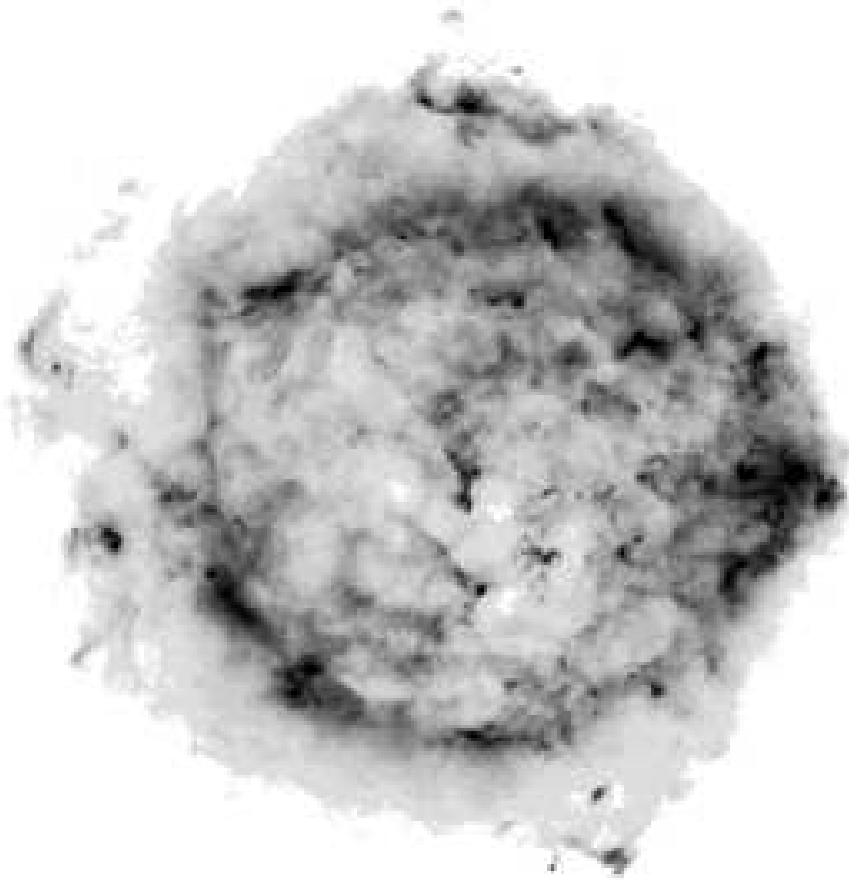}

\caption{Cas~A observed in 6 cm radio wavelengths with the VLA (R.J. Tuffs, 1986).  
The bright ring of emission is interpreted as the result of a compressed
circumstellar wind shell [51], with strongly amplified magnetic field [53].
\label{figure2}}

\end{figure}

\section{Test of the Supernova Remnant origin of Cosmic Rays}

\noindent Following the hypothesis of Baade and Zwicky from 1934, that Supernova
explosions might be the ultimate sources of the CRs [1], the idea has been
entertained in various forms. Indeed as time progressed it became more and more
clear that there are not many viable alternatives from an energetics point of
view. And it has slowly become a common belief that the shocks in the
circumstellar medium, produced by the violently expanding SNRs, should accelerate
the CRs in the Galaxy up to about the knee in the all-particle energy spectrum at
several $10^{15}$~eV.

From an experimental point of view it is obvious that any real test can only 
come through characteristic radiation signatures from the dominant nuclear
particles in individual objects, at comparable energies. The most direct
way to achieve this test is the detection of gamma-rays at TeV energies.

\subsection{Practical beginning of the source search}

\noindent This effort began in earnest about ten years ago when time-dependent
nonlinear acceleration models in SNRs, still in a hydrodynamic approximation also
for the CR component, were used to calculate the expected \gr emission. The
models had to assume the form of the energetic particle spectrum, consistent with
the calculated time evolution of the total energy in CRs, and in detail they turned
out to be sensitive to the assumed injection rates at suprathermal energies [2,3].
While these dynamic models concentrated on the nuclear \grs from $\pi^0$-decay
following inelastic collisions with nuclei from the thermal gas, there was also a
class of kinematic models by e.g. Naito \& Takahara [4] and Gaisser et al. [5] who
rather assumed distributions of accelerated nuclei and energetic electrons but then
concentrated on the various radiation processes from these particle populations and
their relative importance. Subsequently \gr emission models were developed from
kinetic theory by Berezhko \& V\"olk [6,7]. They involved numerical solutions of
the full time-dependent, nonlinear CR transport equation in spherical symmetry
which had been first obtained by Berezhko et al. [8,9]. Based on stationary plane
wave solutions of the same transport equations with a Monte Carlo code [10], Baring
et al. [11] also estimated the time-dependent \gr emission.

The conclusions from this theoretical work were cautiously optimistic: nuclear
TeV \grs from bright nearby objects should be marginally detectable by existing
ground-based instruments.

\subsection{Early observational attempts}

\noindent The first specific \gr observations were made in the early nineties.  
However the initial results were rather inconclusive. Esposito et al. [12] had
observed several shell-type SNRs, notably G78.2+2.1 ($\gamma$-Cygni) and IC443,
with the EGRET instrument on CGRO at \gr energies below 1 GeV. Subsequent TeV
observations by the Whipple [13] and HEGRA [14] groups yielded only upper limits.
For $\gamma$-Cygni and IC 443 these upper limits were below the power-law
extrapolations of the EGRET spectra and, like the Whipple upper limit for Tycho's
SNR, they tended also to lie below the theoretical estimates [3] for the
$\pi^0$-decay \gr flux. In contrast a first TeV-detection was reported for the
Southern Hemisphere remnant of SN~1006 by the CANGAROO collaboration [15]. The flux
value given significantly exceeded the expected $\pi^0$-decay flux [16].

The failure to detect TeV emission from the EGRET sources could be understood if
in reality a straight extrapolation of the power law spectrum from EGRET energies
to the TeV range was not required. The expected emissions would then come from
different objects. Alternatively the remnant could be "old"  in an evolutionary
sense, having lost many of its very high energy particles already (see the
section on Cas~A). Such doubts were substantiated by Brazier et al. [17] who
rather found a Pulsar in the EGRET 95 \% error box for $\gamma$-Cygni, and
Pulsars are generally assumed to exhibit a cutoff in the emission spectrum at
some tens of GeV. In IC443 Keohane et al. [18] found only two regions with hard
X-ray emission from high energy ($\sim 10$~TeV) electrons, certainly not over
most of the radio shell. An association between the hard X-ray sources and the
EGRET source can not be firmly established either (see [19], which also contains
a summary of the recent X-ray results). More generally speaking, no unambiguous
EGRET detection of a shell-type SNR exists. This is one of the questions which
the GLAST mission will have to resolve.


\subsection{SNR detections in nonthermal hard X-rays and TeV \grs}

\noindent In the mid-nineties SN~1006 was detected in hard X-rays [20,21] and
much of this emission was attributed to synchrotron radiation of electrons with
energies of tens of TeV [20,22], see Fig. 3a. Subsequently also several other
shell-type remnants where shown to have power law tails in their emission,
notably SNR RX J1713.7-3946 in the Southern Hemisphere [24,25] and Cas~A [26].
 
\begin{figure}[htbp]

\centering

\includegraphics[width=0.45\linewidth]{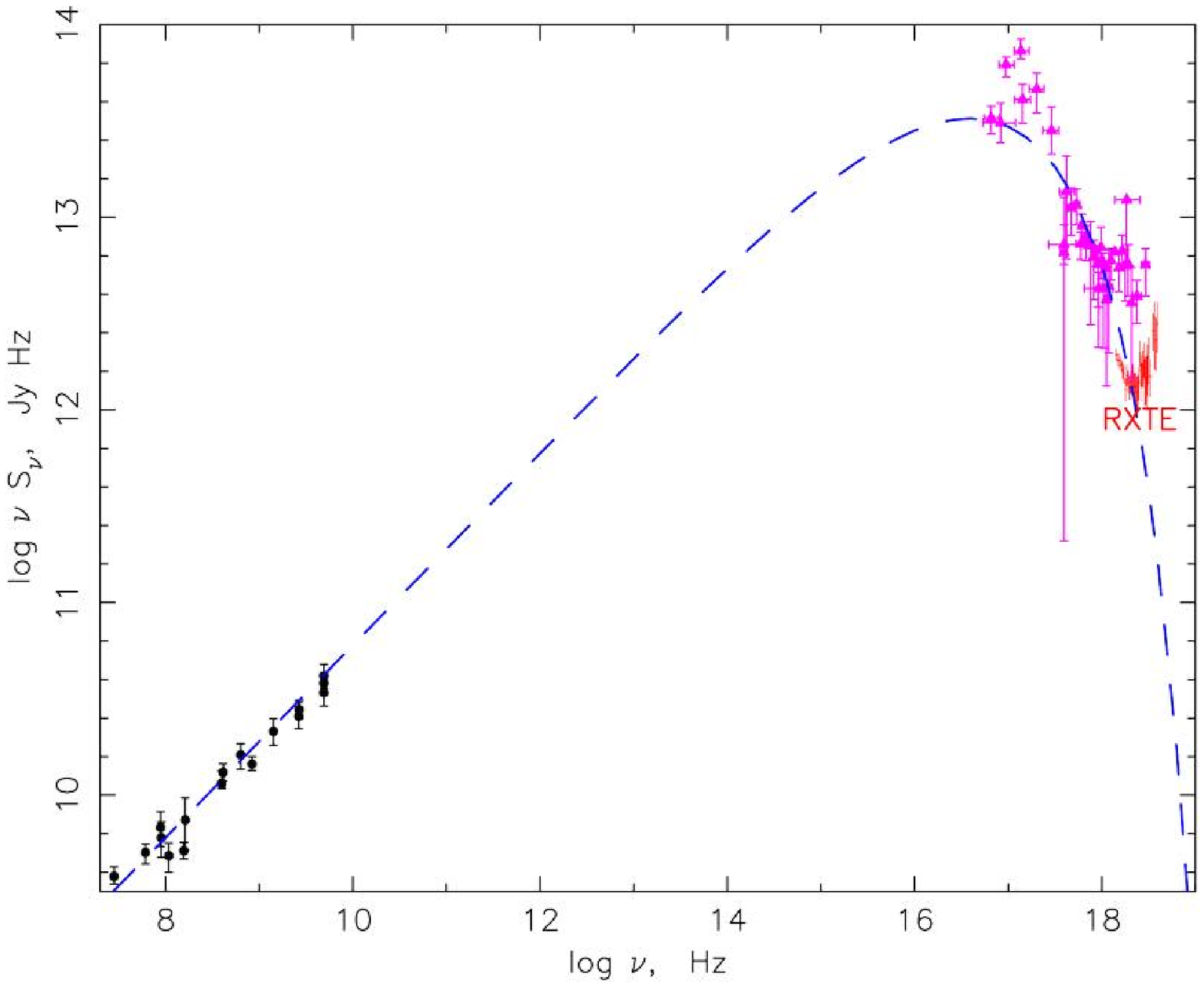}\hspace{0.4cm}
\includegraphics[width=0.45\linewidth]{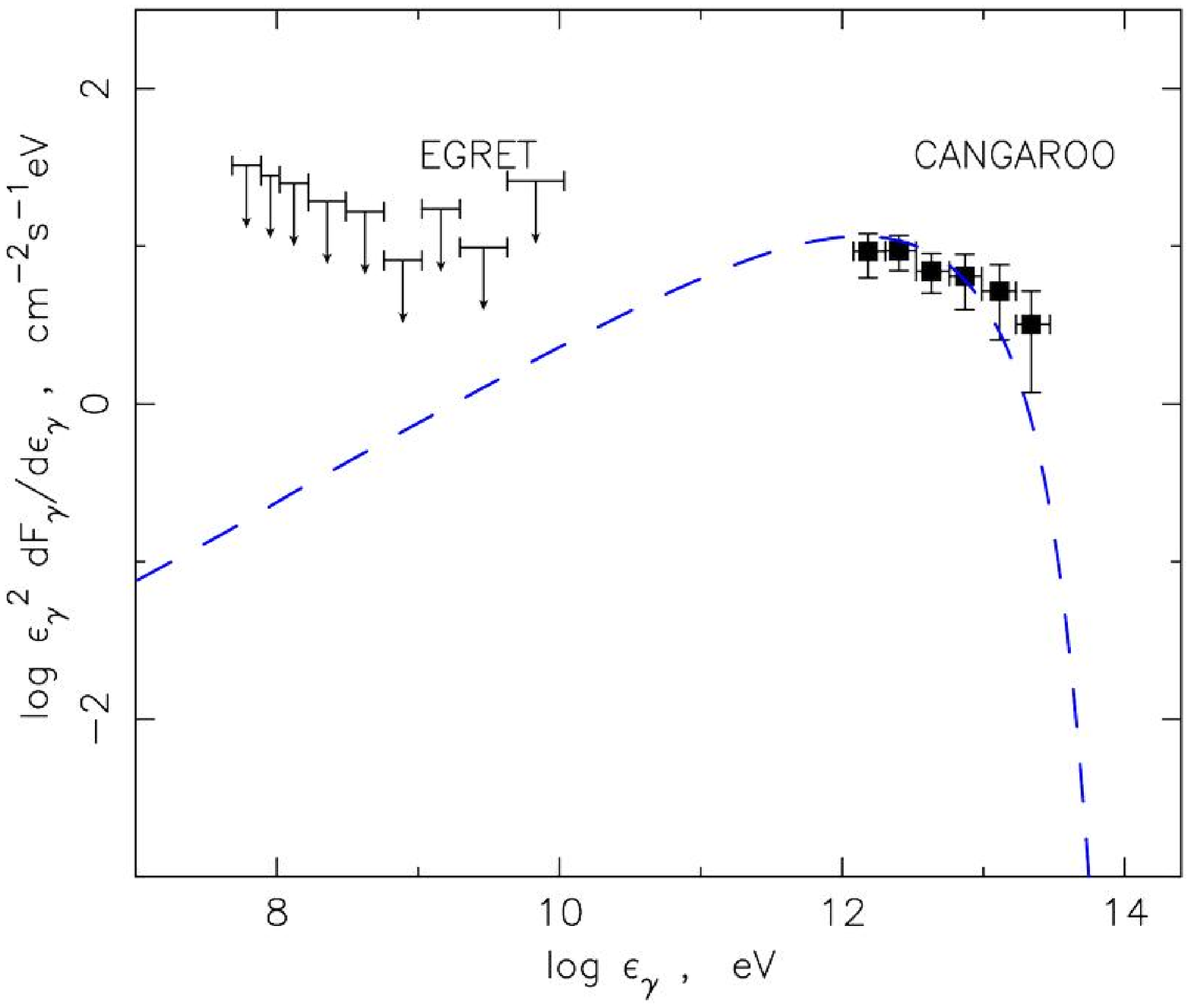}

\caption{a. Synchrotron and \gr emission from SN~1006. Radio data are from
[22], X-ray data are from [21] and [23]. Approximately a power law plus
cutoff ({\it dashed curve}) for the electron distribution in a weak field
of several $\mu$G has been used in a phenomenological fit of the data. b.
The same electron spectra are used for an IC fit to the differential \gr
energy flux data reported in [36]. Also the EGRET upper limits are shown.  
\label{figure3}}

\end{figure}

The question was then, whether the energetic electrons would not also be visible
in TeV \grs through their Inverse Compton (IC) scattering on ambient low energy
photons, primarily from the Cosmic Microwave Background (CMB) [27,28]. In 1998
the CANGAROO collaboration published its detection of SN~1006 [15] mentioned
earlier, and interpreted it in terms of IC emission. A subsequent TeV detection
of SNR RX J1713.7-3946 with the same telescope [29] was given an analogous
interpretation. The only critical discussion at the time was published by Atoyan
and Aharonian [30].

The prevalent opinion that there was no need for nuclear CRs to explain the TeV
\gr emission from SNRs gave room to doubts as to the existence of any significant
quantities of CR nuclei in shell-type SNRs at all [31]. However this pessimistic
turn was provoked by phenomenological arguments not by theory, as all the
arguments before. Again on a phenomenological basis, the CANGAROO collaboration
on the other hand reversed its view regarding SNR RX J1713.7-3946 by presenting
new arguments which now favored a hadronic \gr origin [32] (see also [25] above).
This led to a controversial discussion, where several groups [33,34] gave
empirical counterarguments which questioned the new interpretation. Unfortunately
SNR RX J1713.7-3946 is a complex source which may either be the result of the
thermonuclear explosion/deflagration of an accreting White Dwarf (SN Type Ia)  
that ejects a Chandrasekhar mass, or of the core collapse of a massive star. The
distance uncertainty ranges from 1 to 6 kpc [24, 25]. The remnant is situated in
a complex interstellar environment (e.g. [35]) and is far less well studied than
SN~1006, for example. Therefore the debate remains quite inconclusive at present.
SN~1006 is a much clearer case. It is a SN type Ia, with excellent radio and
X-ray observations and extensive morphological studies. Fig. 3b shows the latest
published \gr results [36], with an IC fit for the spectrum.

Our own conclusion was that phenomenological considerations are important
but that they will not give a convincing answer to the question of the
acceleration of CR nuclei in SNRs as reasonable as they may appear
individually, one by one. There is a need for quantitative comparison
of the observations with a consistent theory, as in other areas of
physics. Contradictory interpretations of the observations can be excluded
only if such a detailed picture is available.

\section{Theory of diffusive shock acceleration as applied to SNRs}

\noindent Let me summarize such a theory in three cartoons. Obviously they are only
meant to illustrate the essential physics; the real description is given by the
equations which I shall only write down to hint at the mathematical aspects. Much
of the theory is reviewed in e.g. [37,38,39,40,41].
\begin{figure}[htbp]
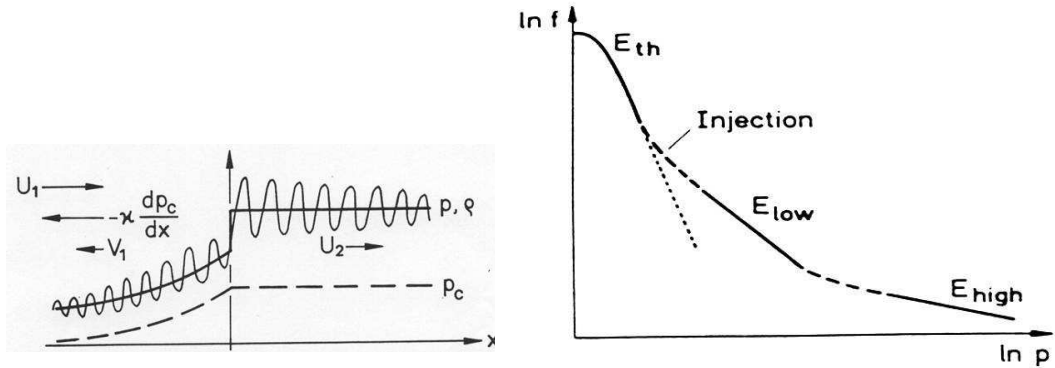


\centering

\includegraphics[width=0.49\linewidth]{Bild1_test.epsi}\hspace*{-0.5cm}
\includegraphics[width=0.49\linewidth]{Bild2.epsi}

\caption{Spatial dependence of flow velocity $U_i$, gas pressure (denoted here by
$p$), mass density $\rho$, and CR pressure $p_c$, in the regions upstream (i=1)
and downstream (i=2) of a plane shock, as a function of the space coordinate $x$
in the shock frame of reference. The wavy line indicates the growing amplitude of
the scattering magnetic fluctuations of the Alfv\'{e}nic wave field which is
excited by the diffusive CR current $-\kappa dp_c/dx$ into the upstream precursor
region and compressed in the subshock at $x=0$. $V_1$ is the wave phase velocity
vector in the upstream fluid frame. Particles gain energy by repeated scatterings
across the shock (left panel). The right panel shows the momentum dependence of
the downstream particle distribution. The three regions joined by dashed lines
correspond to the thermal plasma, characterized by the thermal energy ($E_{th}$)
and extrapolated by a dotted curve, the low-energy nonthermal particles
($E_{low}$), and the high-energy particles ($E_{high}$), respectively.  
\label{figure4}}

\end{figure} 

\noindent The left cartoon of Fig. 4 shows the spatial dependence in a plane
shock, propagating parallel to the magnetic field. The nonlinear backreaction of
the accelerated particles on the dynamics of the thermal gas through their
pressure gradient creates a smooth precursor, followed by a sharp subshock, where
the fluid quantities of the thermal gas jump discontinuously. The CR pressure

\begin{equation}
p_c={4\pi c\over 3} \int_{p_{inj}}^{\infty}dp{p^4f(x,p,t)\over\sqrt{p^2+m^2c^2}}
\end{equation}

\noindent is given in terms of the isotropic part of the particle distribution
function $f(x,p,t)$, averaged over the magnetic fluctuations; the quantities
$p$ and $m$ denote the magnitude of particle momentum and its mass. In the
downstream distribution (right panel) CRs are characterized by momenta above the
region of injection where downstream particles can outrun the shock along the
magnetic field. The nonthermal spectrum starts with a steep power law
distribution, characterized by diffusive acceleration at the subshock only. High
energy particles "see" the entire compression as a discontinuity and have
therefore a harder spectrum that begins in the region where protons become
ultrarelativistic [9].

Strong shocks with Alfv\'enic Mach numbers $M_a = U_1/V_1 \gg 1$ typically inject
so many suprathermal nuclear particles into the acceleration process that the
acceleration is efficient, $p_c \sim \rho_1 U_1^2$, and that nonlinear
backreaction leads to a strong precursor. In SNRs $M_a$ is of order $10^3$ at
very early times, and for SN~1006 it is still as large as 150 at the present
epoch. Thus the total compression ratio $r_{tot} > 4$, whereas the subshock
compression ratio $r_{sub} < 4$. Only part of the overall energy dissipation goes
into thermal energy at the subshock, the rest is given to the energetic particles
and to the growing waves.

The most difficult aspect of particle acceleration is the magnetic field and its
fluctuations, because this field is what the energetic particles interact with
directly. And the field is not simply given externally, but is rather excited by
the particles themselves as another part of the nonlinearity of the process.

The selfconsistent wave production by the accelerated particles themselves can
only be estimated rather approximately. A simple-minded quasilinear calculation
[42] for efficient acceleration leads to an extremely large wave field $\delta
B/B \gg 1$, where $\delta B$ is the total rms wave amplitude and $B$ is the
assumed mean magnetic field strength. This conclusion has been shown to hold also
differentially for the wave energy per unit logarithmic bandwidth in wave number
in relation to the resonant energetic particle pressure $P_c(p) = 4\pi c/3~
p^4f(x,p,t)/\sqrt{p^2+m^2c^2}$ per logarithmic interval in $p$ [41].

However, such a result defeats the original assumption $\delta B/B \ll 1$ of
perturbation theory, on which the derivation of the result is predicated (I will
loosely continue to use the term $\delta B$ for both the differential and the
integral wave amplitude). The question is then how to deal with this outrageous
situation. In fact, theory has no precise solution at the moment. Apart from wave
damping, the simplest and physically most plausible conclusion is that the strong
wave production amplifies the mean field to an effective field $B_{\mathrm eff}
\gg B_0$ [38], with $\delta B/B_{\mathrm eff} \sim 1$. In this limit the
scattering mean free path $\lambda_{mfp}$ reaches its minimum value
$r_{gyro}(p)$, where $r_{gyro}$ is the particle gyro radius in the field
$B_{\mathrm eff}$. This is the so-called Bohm limit in the effective field and
implies an isotropic diffusion particle coefficient $\kappa(p,B_{\mathrm eff}) =
v r_g(p)/3$. Field amplification has recently been successfully calculated in a
simplified nonlinear model of wave turbulence by Lucek \& Bell [43] and Bell \&
Lucek [44]. The result is the Bohm limit. As a scaling relation for strong shocks
it is already suggested by perturbation theory.

Finally strong wave production has also profound consequences for the rate and
the geometry of ion injection (third cartoon of Fig. 5a): where on the shock
surface injection can occur in the first place -- in an approximately spherical
shock propagating in an inhomogeneous external magnetic field this is only
possible in the "polar"  regions -- it occurs only at those field lines that are
instantaneously quasi-parallel to the subshock normal direction. On all other
field lines injection is suppressed until the field changes back to
quasi-parallel. This has the consequence that the ion injection rate (the
fraction of the incoming flux of thermal particles that can be accelerated) is
reduced by about two orders of magnitude compared to that in a laminar magnetic
field. The precise numerical value of the reduction factor depends on the
wave spectrum. For a uniform ambient field the systematic reduction of injection
will lead to dipolar asymetries in the nonthermal morphology and requires a
renormalization of the overall particle production rate, as calculated in
spherical symmetry, by a factor $f_{re} <1$ [45].

\begin{figure}[htbp]

\centering

\includegraphics[width=0.49\linewidth]{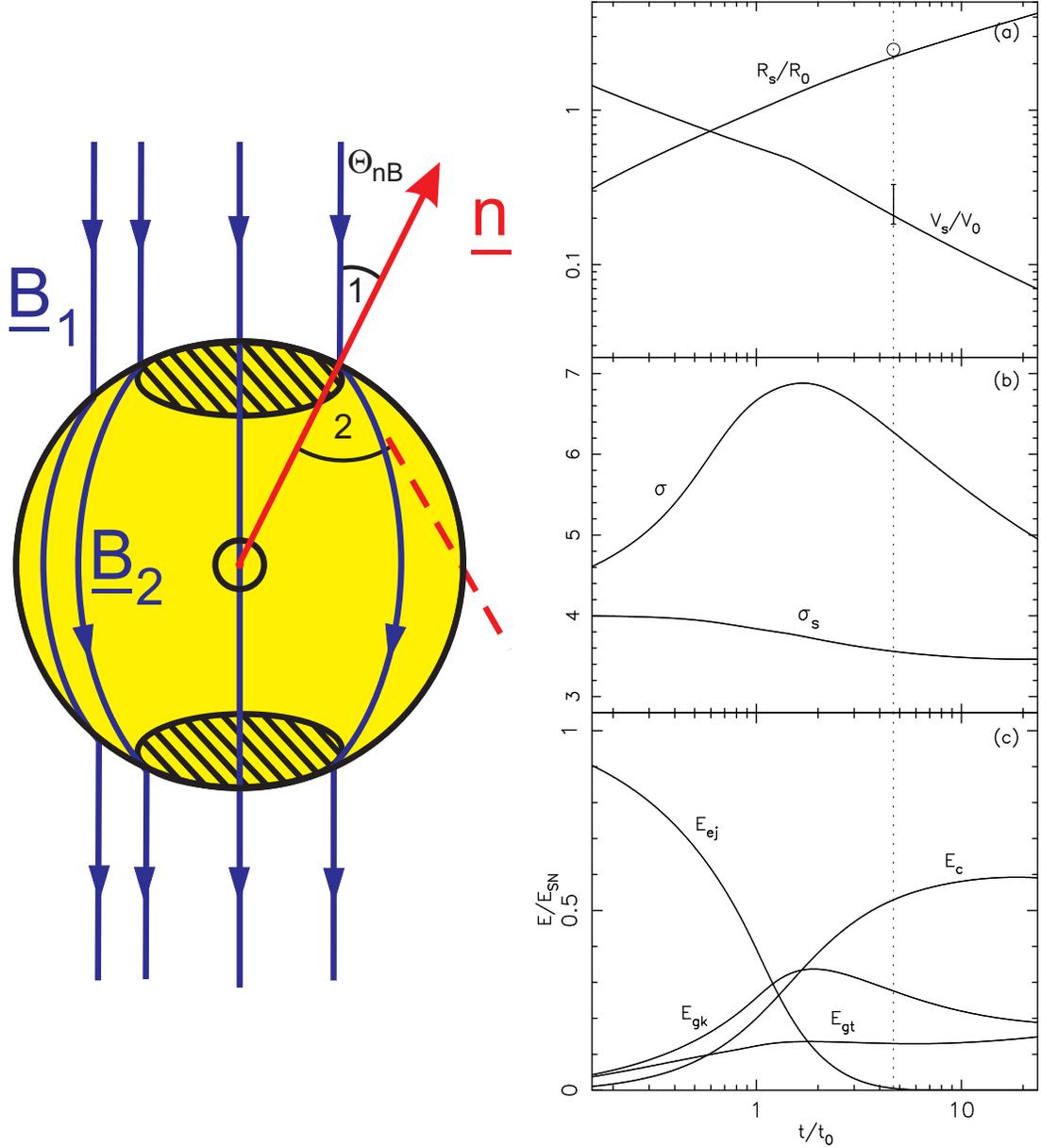}\hspace*{2mm}
\includegraphics[width=0.49\linewidth]{SN1006_fig1.eps}

\caption{Left panel: Ion injection into diffusive shock acceleration. Only for
$\Theta_{nB_2}$ sufficiently smaller than $90^{\circ}$ suprathermal ions can
escape upstream along the magnetic field $\underline B$. In a spherical SNR in a
uniform field $\underline {B}_1$, injection, strong wave production and
acceleration is only possible in the {\it hatched} polar regions. Therefore
hadronic \gr emission is dipolar and the same is true for the synchrotron
emission as a result of field amplification. The right panel contains the
temporal evolution of the overall dynamics for SN~1006 [46]. For external density
$n=0.3~{\mathrm cm}^{-3}$ the shock radius $R_s$ and velocity $V_s$ are exhibited
together with the observed values. Total, $\sigma$, and subshock, $\sigma_s$,
compression ratios are given in the middle panel. The quantities $E_{SN}$,
$E_{ej}$, $E_{gk}$, $E_{gt}$ and $E_c$ denote the total SNR mechanical, ejected,
gas kinetic, gas thermal, and un-renormalized CR energy contents, respectively.  
\label{figure5}}

\end{figure}

The resulting theory, applied to the dynamics of a SNR and assuming spherical 
symmetry, looks as follows:

CR transport and acceleration at shocks is described by a Fokker-Planck
equation for the isotropic part of the mean particle distribution 
$f_{i}(p,r,t)$ for ions (p) and electrons (e), respectively:

\begin{equation}
{\partial f_\mathrm{i}\over \partial t}
-\nabla \kappa \nabla
f_\mathrm{i}
+\vec{U}\nabla f_\mathrm{i}
-\frac{\nabla \vec{U}}{3}p\frac{\partial f_\mathrm{i}}{\partial p}
+\frac{1}{p^2}\frac{\partial}{\partial p}
\left( \frac{p^3}{\tau_\mathrm{i} }f_\mathrm{i}\right)= Q_i, 
\label{eq3}
\end{equation}

\noindent where the coefficients in the transport eqs. are given by the mean mass
velocity $\vec{U}(r,t)$ and its divergence, as determined from the hydrodynamic
eqs., and by a common diffusion coefficient $\kappa (p, B_{\mathrm eff})$ as a 
result of
scattering by the fluctuating magnetic field. $Q_i$ denotes the injection rate,
and $\tau_\mathrm{i}$ is the loss time, taken to be zero for ions and equal to
the synchrotron loss time for electrons. 

The hydrodynamics of the thermal plasma, even if we consider it as an ideal
fluid, is in turn nonlinearly coupled with the energetic particles through the CR
pressure gradient $dp_c/dx$, defined above, as well as wave dissipation, see e.g. 
[7]:

\begin{equation}
{\partial\rho \over \partial t}+ \nabla(\rho \vec U)=0,
\end{equation}

\begin{equation}
\rho{\partial \vec U \over \partial t}+\rho (\vec U\nabla)\vec U=
-\nabla(p_c +p_g),
\end{equation}
 
\begin{equation}
{\partial p_g \over \partial t}+(\vec U \nabla)p_g+\gamma_g
(\nabla\vec U)p_g=\alpha_a(1-\gamma_g)c_a\nabla {p_c},
\end{equation}

\noindent 
where $\rho$, $\gamma_g=5/3$ and $p_g$ denote the mass density, gas specific
heat ratio and gas pressure, respectively. Gas heating due to Alfv\'en wave
dissipation in the upstream region is described by the parameter $\alpha_a \leq
1.$

Already in this lowest approximation the theory involves a time-dependent
nonlinear system of coupled partial integro-differential equations.

In principle the theory is quite general and can be used to describe any ionized
system in which matter dominates and thermal gas motions are nonrelativistic. The
above application to point explosions like SNRs is the simplest one from an
analytical point of view. Astrophysically speaking it is nevertheless
fundamental. In this case the theory contains only two "parameters" that are not
very well calculable quantitatively: $B_{\mathrm eff}$, and the ion injection
rate $Q_p$. They are strongly related through the nonlinear development of wave
production discussed above. It is therefore plausible to use an
experimental/observational input for these parameters in order to be free from
uncertain theoretical approximations.

How do we achieve this observational input for SNRs? The natural solution is
given in terms of the electron component and the corresponding sychrotron
observations [46]. The electrons are parasitically accelerated together with the
nuclear particles which generate the scattering wave field through their dominant
mass and energy density, at least above the (unknown) electron injection energy.
This means that above injection the overall electron momentum distribution must
have equal form to that of the nuclear particles except for radiative losses. The
synchrotron observations then yield three quantities: (i) the radio synchrotron
spectrum will be steeper than in the test particle approximation; interpreting
this in terms of nonlinear modification determines then the ion injection rate
(ii) the requirement that these radio electrons have energies $\leq 1 $~GeV gives
the value of $B_{\mathrm eff}$ (iii) the electron density amplitude from the
magnitude of the synchrotron emission determines the electron:proton ratio. The
actual average nuclear CR density in the SNR is reduced from the calculated
spherically symmetric ion amplitude by the ratio $f_{re}$ of the area of
efficient injection to the total SNR surface area (Renormalization)[45].

\section{SN~1006}

\noindent In the radio and in nonthermal X-rays SN~1006 has a dipolar structure
[20,47]. This is consistent with ion injection theory, as described above and
with a diameter of $0.5^{\circ}$ it should be an extended TeV source as well,
allowing a test of the \gr morphology. From astronomical measurements at radio
and X-ray wavelengths the distance is about 1.8 kpc, and the expansion velocity
amounts to 3000 km/sec. A rather uncertain astronomical parameter is the density
$n$ of the ambient Interstellar Medium (ISM). In the following we use a value $n
=0.3~{\mathrm cm}^{-3}$ which is suggested by X-ray measurements. However, the
density could be as low as $n =0.1~{\mathrm cm}^{-3}$ since SN~1006 is located
relatively far above the Galactic Plane. This also suggests a standard
interstellar magnetic field of several $\mu$G, rather uniform on the scale of the
SNR. However, as emphasized below, the effective field is expected to be
substantially higher than this.

\subsection{Model calculations for SN~1006}

\noindent I shall now paraphrase the detailed model calculations [46] that are
based on the theory discussed before. The overall hydrodynamic quantities are
given in Fig. 5b. From a simultaneous fit of the observed SNR radius and
expansion speed at the known age of the system of almost 1000 years the total
mechanical explosion energy must be chosen as $E_{SN} = 3 ~10^{51}$; $E_{SN}$ is
somewhat high compared to the canonical $10^{51}$~erg. The total shock
compression ratio is substantially above the adiabatic value for a strong shock.
The spherically symmetric calculation gives a total CR energy $E_c \approx 0.6
E_{SN}$ at late times. However, the renormalization factor is 0.2 [45]. This
results in a fraction of about 10 \% in nuclear CR energy at the late Sedov
phase, when the source particles are expected to be released into the ISM.

\begin{figure}[htbp]
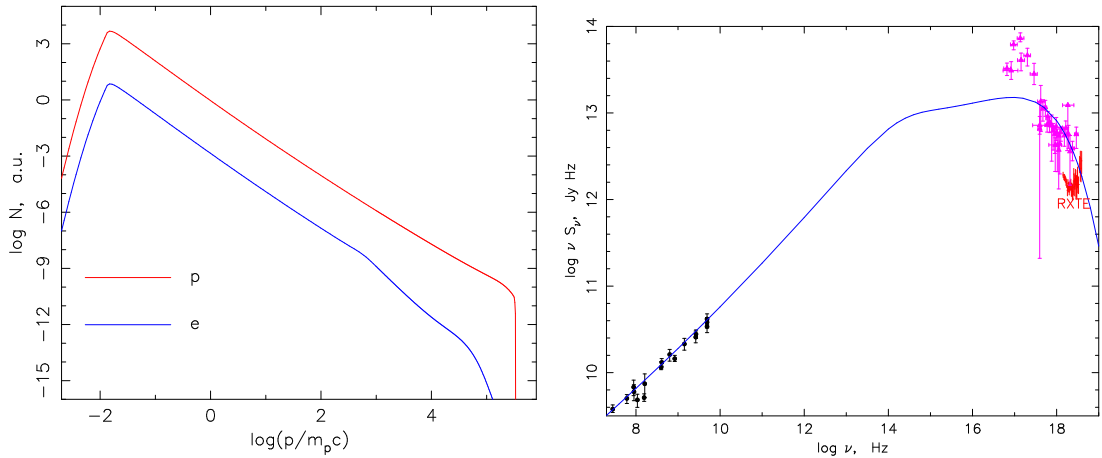


\centering

\includegraphics[width=0.485\linewidth]{SN1006_sn2.eps}\hspace*{2mm}
\includegraphics[width=0.49\linewidth]{SN1006_fig3_high_field.eps}

\caption{Volume-integrated particle distributions $N$ and synchrotron spectra
$\nu S_{\nu}$ for accelerated protons (p) and electrons (e) in SN~1006 [46].  
The observational data are the same as in Fig. 3a. 
\label{figure6}}

\end{figure}

\noindent The spatially integrated spectra of CR electrons and nuclei(Fig. 6a)
harden towards the cutoff, except for electron cooling. The high effective
internal field $B_{\mathrm eff} \approx 120~\mu$G is determined from the spectral
slope of the radio synchrotron spectrum in Fig. 6b which also fixes the injection
rate to about $10^{-4}$ times the flux of incoming particles. The flattening of
the high frequency synchrotron spectrum also allows a smooth connection below the
thermal X-rays to the hard X-ray data points. All this implies a somewhat low
predicted electron:proton ratio in the overall number of accelerated particles of
about $2 \cdot 10^{-3}$ for this source, compared to the canonical interstellar
value of $10^{-2}$.

The predicted contributions to the differential \gr energy spectrum, integrated
over the SNR volume, are shown in Fig. 7a and the radial \gr brightness at 3 TeV
is exhibited in Fig. 7b. For the assumed ISM density $n$ the IC contribution from
CR electrons is about one order of magnitude below the hadronic contribution due
to $\pi^0$-decay, which in turn is still below the EGRET upper limits. As
consequence of synchrotron cooling, the IC emission spectrum has an almost
identical form to the hadronic \gr spectrum from several GeV up to the cutoff.
This makes attempts for a distiction of spectral slopes very difficult except at
very low and very high energies.

It is in this context important to note that the $\pi^0$-decay \gr flux scales
like $n^2$ to lowest order, although the strong nonlinearity of the system makes
this scaling only a rough first approximation. Decreasing the external density by
a factor of 2 will therefore diminish the hadronic \gr flux by a factor of about
4. At the same time the electron:proton ratio will increase, whereas the derived
total energy $E_{SN}$ will decrease. An external density which is lower than $n
=0.3~{\mathrm cm}^{-3}$ may explain the fact that the individual H.E.S.S.
telescopes in Namibia have not been able to detect SN~1006 at the level suggested
by the CANGAROO results.

Fig. 3b in fact indicates an IC spectrum similar to the one used to
explain the more recent data points given by the CANGAROO II telescope
[36]. Such a spectrum is the result of a scenario that neglects
acceleration of nuclear particles and assumes an average internal field of
about $10~\mu$G, appropriate for an MHD-compressed ISM magnetic field of
several $\mu$G. The form of this spectrum is different from both the
hadronic and the cooled IC spectra above a few GeV.  Nevertheless, a
well-defined observational \gr spectrum is required to distinguish this
scenario from the solid line spectra calculated from the nonlinear theory
if the fluxes are comparable.

\begin{figure}[htbp]
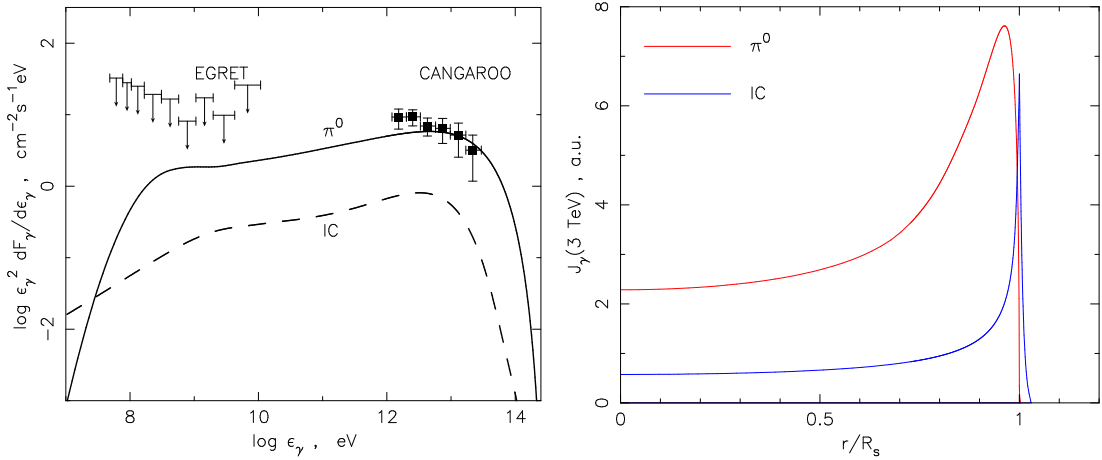


\centering

\includegraphics[width=0.49\linewidth]{SN1006_fig5_snrgr.ps}\hspace*{2mm}
\includegraphics[width=0.49\linewidth]{SN1006_sn7.ps}

\caption{Predicted differential energy spectrum of $\pi^0$-decay and IC
emission (left panel), and \gr brightness (at 3 TeV) as functions of radius $r$
in units of the shock radius $R_s$ (right panel), for SN~1006 for an ambient
density $n=0.3~{\mathrm cm}^{-3}$. Upper limits from EGRET and the reported
CANGAROO fluxes are indicated [46]. 
\label{figure7}}

\end{figure}

The picture becomes richer when we consider the morphology as well, for instance
the radial dependence in the neighborhood of the dipole axis (Fig. 7b, for a \gr
energy of 3 TeV). Here the IC emission is concentrated into an extremely small
scale $l_e \sim 10^{-2} R_s$ near the shock at $r=R_s$ as a consequence of
synchrotron cooling of the generating $\sim 100$~TeV electrons. Also the hadronic
component is fairly strongly confined to a narrow shell, since the thermal gas is
concentrated there. The emission morphology for a low-field scenario (not shown
here) would be considerably more extended, only exhibiting the large-scale
adiabatic expansion losses in the interior. In summary, the \gr emission is
expected to show a dipolar spatial morphology [48], similar to that of the
synchrotron emission.

An important observational result has been recently been found in Chandra
observations at X-ray energies in the 1 to 10 keV region [47,49] (Fig. 8a). 

\begin{figure}[htbp]
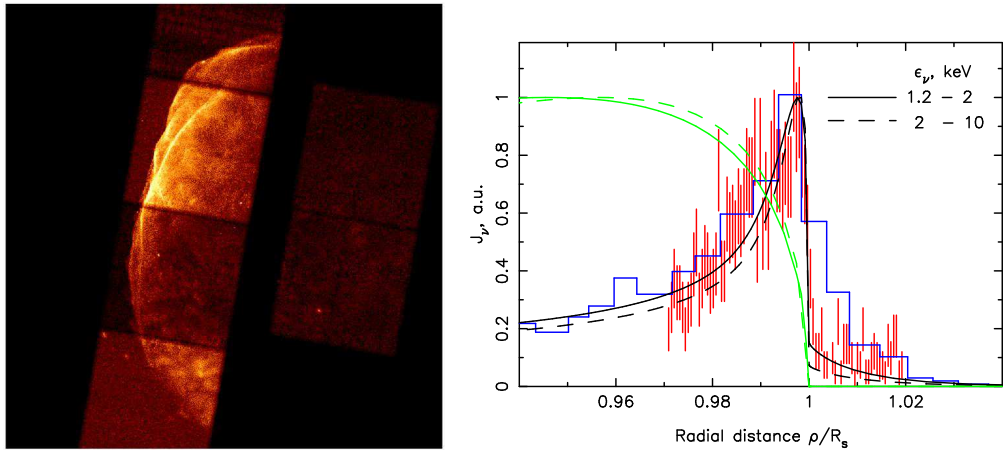


\centering

\includegraphics[width=0.4\linewidth]{Bild5.epsi}\hspace*{2mm}
\includegraphics[width=0.49\linewidth]{xscale_fig2.eps}

\caption{Northeast limb of SN~1006, as seen by Chandra between 0.5 and 2
keV [47] (left), and projected X-ray brightness of the most narrow
structure, compared with the theoretical prediction [50] (right). Vertical
dashes are from [49], histogram from [47]. The thin curves correspond to
the low-field scenario in Fig. 3a.  
\label{figure8}}

\end{figure}

\noindent They demonstrate very narrow spatial X-ray emission structures, on a
scale of $10^{-2} R_s$, confirming the magnetic field amplification made possible
by a nonthermal particle population dominated by nuclear particles [46,50]. In
contrast, the low-field scenario clearly fails when compared to the observations
(Fig. 8b).

As a corollary, the electron population cannot create the strong field
fluctuations and the required high effective field by itself [50]. This means
that the dominant accelerating nuclear component is not only possible
experimentally as well as theoretically: it is also necessary.

\section{Cassiopeia~A}

\noindent This SN type Ib, whose progenitor was probably a massive Wolf-Rayet
star, is on the other end of the scale of Supernovae when compared with SN~1006.
Such massive progenitor stars modify the circumstellar medium through substantial
mass-loss during several successive stellar wind phases: the fast rarefied Blue
Supergiant wind during the main sequence phase turns into a massive but slow Red
Supergiant (RSG) wind at a late stage, to be subsequently compressed from inside
by yet another fast wind, now from the emerging Wolf-Rayet star, until this star
finally collapses as Supernova. The turbulent compressed shell of the RSG
material has a high gas density of about $10~\mathrm {cm}^{-3}$ owing to
radiative cooling and is identified with the so-called bright ring in Cas A [51],
see Fig. 2. According to this picture, the SNR's leading shock has already
reached the unperturbed RSG wind region.

\begin{figure}[htbp]

\centering

\includegraphics[width=0.49\linewidth]{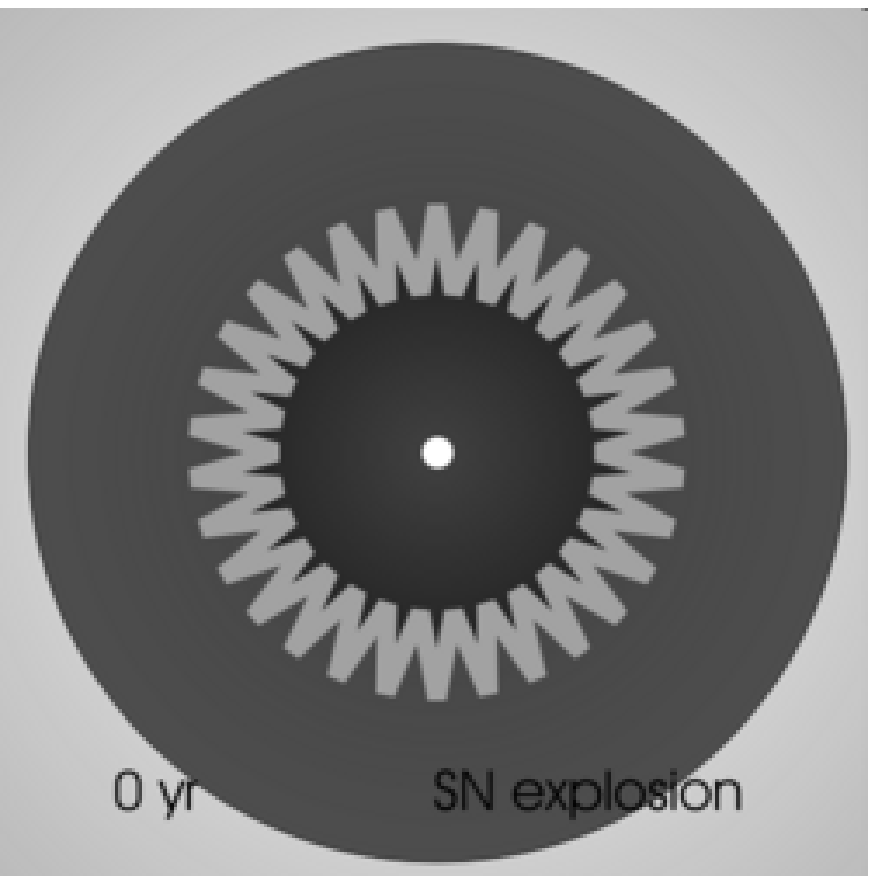}\hspace*{2mm}
\includegraphics[width=0.49\linewidth]{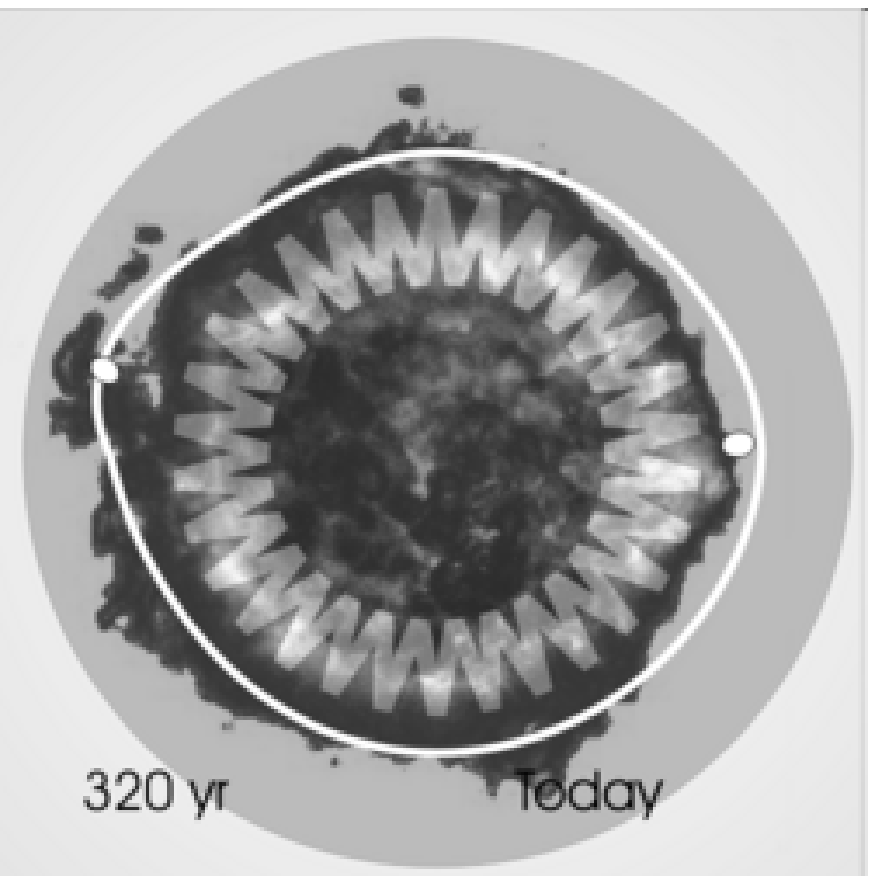}

\caption{Schematic of Cas~A's circumstellar environment at different stages of
evolution.  Left panel: At explosion the turbulent shell is bordered by the
interior Wolf-Rayet wind bubble and the exterior Red Supergiant wind region.
Right panel: Today's SNR shock and clumpy ejecta (white spots) beyond the
shocked shell and part of the Red Supergiant wind region, superposed on radio
image of Fig. 2. (Courtesy G. P\"uhlhofer). \label{figure9}}

\end{figure}

The HEGRA telescope system observed Cas~A for 232 hours, the longest
pointing used in \gr astronomy until now, and finally detected the source at
3.3 percent of the Crab level [52]. The acceleration model summarized here [53]
basically follows the picture described above and assumes a Parker spiral type
mean circumstellar magnetic field topology. The initial configuration before
explosion is schematically shown in Fig. 9a, while Fig. 9b pictures the
situation of today.

\begin{figure}[htbp]
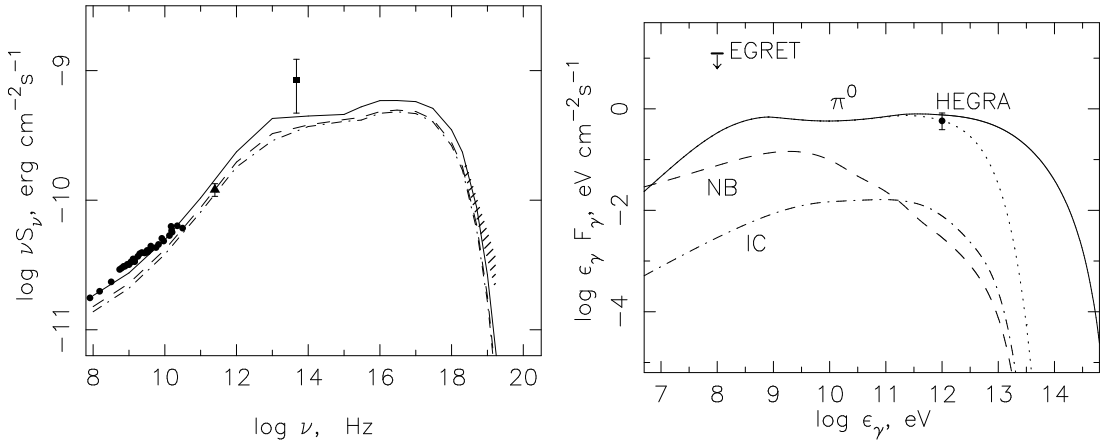


\centering

\includegraphics[width=0.49\linewidth]{Casa_fig3n.eps}\hspace*{2mm}
\includegraphics[width=0.49\linewidth]{CASA_fig5n.eps}

\caption{Calculated synchrotron spectra at epochs 1970 (\it solid curve), 
2002 (\it dashed curve) and 2022 (\it dashed-dotted curve), respectively,
together with observational data over a comparable time span (left panel). The
right panel shows the predicted integral \gr energy spectrum: Nonthermal
Bremsstrahlung (NB, {\it dashed line}) and IC emission ({\it dash-dotted line}),
together with dominant $\pi^0$-decay emission ({\it solid line}) are far below the
EGRET upper limit, but agree with the HEGRA flux at 1 TeV. The {\it dotted line}
indicates a possibly lowered proton cutoff as a result of particle escape from the
SNR [53].
\label{figure10}}

\end{figure}

Given the synchrotron spectrum, the evolution of this complex system containing
the turbulent wind shell allows a consistent description of the SNR dynamics with
a total mechanical energy $E_{SN}=4\cdot 10^{50}$~erg, a present shock speed $V_s
= 2000$~km/sec, and a very large downstream effective field $B_{\mathrm eff} = 1
$~mG in the shell. Even though the available synchtrotron observations are
seperated by 30 years, about 10 \% of the remnant's age, a good fit to the
hardening radio spectrum is possible, followed by a synchrotron cooling range in
the infrared to X-ray range, ending in a cutoff at nonthermal hard X-rays (Fig.
10a). Also the observed temporal decrease of the synchrotron emission is
consistent with the picture of shock propagation into an expanding wind region of
decreasing gas density.

The predicted differential \gr energy spectrum is shown in Fig. 10b. As a
consequence of the high values of gas density and effective magnetic field
strength, it is completely dominated by hadronic \grs from $\pi^0$-decay which
are still two orders of magnitude below the EGRET upper limit. The hadronic
flux has been renormalized by a factor $f_{re} = 1/6$ [45]. This prediction
reasonably agrees with the observed \gr flux value around 1 TeV. The predicted
IC and nonthermal Bremsstrahlung fluxes are very low. We therefore conclude 
that Cas~A is a hadronic \gr source. The predicted proton
spectrum reaches the knee region. Even if escape of CR nuclei from the
weakening shock is likely to set in already at this low age, Cas~A could be
considered as a member of the hypothetical CR source population of Galactic
SNRs.

\section{A few personal theses}

\noindent I want to summarize this talk in a few personal theses. I call them
personal because they may not reflect the position of the two other speakers. On
the other hand, the conclusions are based on the joint efforts of E.G. Berezhko,
L.T. Ksenofontov, G. P\"uhlhofer and myself, and to this extent they are not only
my own conclusions:

\begin{itemize}
\item The astrophysical processes of CR origin up to the knee are basically
understood. The sources are the Galactic SNRs. This concerns 99.9 percent of the
total energy density in CRs.

\item Gamma-ray observations for SNRs are critical since only \grs have
individual energies comparable to those of the generating charged particles.
Currently experimental results are still scarce.

\item The nonthermal synchrotron properties are an indispensible multi-wavelength
aspect of SNRs. This concerns both the morphology and the spectrum.

\item On account of the hard SNR spectra, the most appropriate instruments at \gr
energies are large ground-based detectors like CANGAROO III, H.E.S.S., MAGIC and
VERITAS. They will play a decisive role for the determination of the morphologies
as well as the spectra of individual objects. They are needed to increase the
detection statistics of SNRs.

\item The most important global astronomical test is the detection and spectral
decomposition of the "diffuse" TeV \gr background in the Galactic Plane to
which the entire Galactic SNR population contributes.

\end{itemize}

\section{Acknowledgements}

\noindent I am grateful to F.A. Aharonian, E.G. Berezhko, G. Heinzelmann, L.T.  
Ksenofontov, G.  P\"uhlhofer, and O. Reimer for discussions on various aspects 
of this manuscript.

\section{References}

\re
1.\ Baade W., Zwicky F.\ 1934, Proc. Nat. Acad. Sci. USA, 20, 259
\re
2.\ Dorfi, E.\ 1991, A\&A, 251, 597
\re
3.\ Drury L. O'C., Aharonian F.A., V\"olk H.J.\ 1994, A\&A, 287, 959
\re
4.\ Naito T., Takahara F.\ 1994, J. Phys. G, 20, 477
\re
5.\ Gaisser T.K., Protheroe R.J., Stanev T.\ 1998, ApJ, 492, 219
\re
6.\ Berezhko E.G., V\"olk, H. J.\ 1997, Astropart. Phys., 7, 183
\re
7.\ Berezhko, E.G., V\"olk H.J.\ 2000, A\&A, 357, 283
\re  
8.\ Berezhko E.G., Yelshin V.K., Ksenofontov L.T.\ 1994, Astropart. Phys.,
    2, 215
\re  
9.\ Berezhko E.G., Ksenofontov L.T., Yelshin V.K.\ 1995, Nucl. Phys. B
    (Proc.Suppl.), 39A, 171
\re
10.\ Jones F.C., Ellison D.C.\ 1991, Space Sci. Rev. 58, 259
\re
11.\ Baring M.G., Ellison D.C., Reynolds S.P. et al.\ 1999, ApJ, 513, 311
\re
12.\ Esposito J.A., Hunter S.D., Kanbach G. et al.\ 1996, ApJ, 461, 820
\re
13.\ Buckley J.H., Akerlof C.W., Carter-Lewis D.A. et al.\ 1998, A\&A, 329,
     639
\re
14.\ He\ss~M. (HEGRA Collaboration)\ 1997, Proc. 25th ICRC (Durban), 3, 229
\re
15.\ Tanimori T., Hayami Y., Kamei S. et al.\ 1998, ApJ, 497, L25
\re
16.\ V\"olk H.J.\ 1997, in: de Jager O.C. (ed.) Towards a Major Atmospheric
     Cherenkov Detector -- V, Kruger Park, p. 87 ff.
\re
17.\ Brazier K.T.S., Kanbach G., Carraminana A. et al.\ 1996, MNRAS, 281,
     1033
\re
18.\ Keohane J.W., Petre R., Gotthelf E.V. et al.\ 1997, ApJ, 484, 350
\re
19.\ Sturner S.J., Keohane J.W., Reimer O.\ 2003, in: "High Energy Studies of 
Supernova Remnants and Neutron Stars", 34th COSPAR Symp., Houston 2002, Eds. 
W.Hermsen and W. Becker, Advances in Space Resarch, in press  
\re
20.\ Koyama, K., Petre, R., Gotthelf, E.V. et al.\ 1995, Nature, 378, 255
\re
21.\ Allen G.E., Gotthelf E.V. \& Petre R.\ 1999, Proc. 26the ICRC (Salt
     Lake City), 3, 480
\re
22.\ Reynolds S.P.\ 1996, ApJ, 459 , L13
\re
23.\ Hamilton A.J.S., Sarazin C.L. \& Szymkowiak A.E.\ 1986, ApJ, 300, 698
\re
24.\ Koyama K., Kinugasa K., Matsuzaki K. et al.\ 1007, PASJ, 49, L7
\re
25.\ Slane P., Gaensler B.M., Dame T.M. et al.\ 1999, ApJ, 525, 357
\re
26.\ Allen G.E., Keohane J.W., Gotthelf E.V.\ 1997, ApJ, 487, L97
\re
27.\ Pohl M.\ 1996, A\&A, 307, L57
\re
28.\ Mastichiadis A., de Jager O.C.\ 1996, A\&A, 311, L5
\re
29.\ Muraishi H., Tanimori T., Yanagita S. et al.\ 2000, A\&A, 354, L57
\re
30.\ Aharonian F.A., Atoyan A.M.\ 1999, A\&A, 351,330
\re
31.\ Weekes T.C.\ 2001, in: High Energy Gamma-Ray Astronomy (ed. F.A.
   Aharonian, H.J. V\"olk), AIP Conf. Proc. 558, Melville, New York, p. 15 ff.
\re
32.\ Enomoto R., Tanimori T., Naito T. et al.\ 2002, Nature, 416, 823
\re
33.\ Reimer O., Pohl M.\ 2002, A\&A, 390, L43
\re
34.\ Butt Y.M., Torres D.F., Romero G.E. et al.\ 2002, Nature, 418, 499
\re
35.\ Uchiyama Y., Aharonian F.A., Takahashi T.\ 2003, A\&A, 400, 567
\re
36.\ Tanimori T., Naito T., Yoshida T. et al.\ 2001, Proc. 27th ICRC
     (Hamburg), 6, 2465
\re
37.\ Drury L.O'C.\ 1983, Rep. Prog. Phys., 46, 973
\re
38.\ V\"olk, H. J.\ 1984, in: High Energy Astrophysics, ed. J. Tran Thanh
     Van, Editions Frontieres, Gif sur Yvette
\re
39.\ Blandford R.D., Eichler D.\ 1987, Phys.Rept., 154, 1
\re
40.\ Berezhko E.G., Krymsky G.F.\ 1988, Soviet Phys.-Uspekhi., 12, 155
\re
41.\ Malkov M.A., Drury L.O'C.\ 2001, Rep. Prog. Phys., 64, 429
\re
42.\ McKenzie, J.F., V\"olk, H.J.\ 1982, A\&A, 116, 191
\re
43.\ Lucek S. G., Bell A. R.\ 2000, MNRAS, 314, 65
\re
44.\ Bell A. R., Lucek S. G.\ 2001, MNRAS, 327, 433
\re
45.\ V\"olk H.J., Berezhko E.G., Ksenofontov L.T.\ 2003, A\&A, 409, 563
\re
46.\ Berezhko E.G., Ksenofontov L.T., V\"olk H.J.\ 2002, A\&A, 395, 943
\re
47.\ Long K.S., Reynolds S.P., Raymond J.C. et al.\ 2003, ApJ, 586, 1162
\re
48.\ Konopelko A., Lucarelli F., Lampeitl H., Hofmann W.\ 2002, J. Phys.G,
     28, 2755
\re
49.\ Bamba A., Yamazaki R., Ueno M., Koyama K.\ 2003, ApJ, 589, 827
\re
50.\ Berezhko E.G., Ksenofontov L.T., V\"olk H.J.\ 2003, to appear in A\&A
     Lett. (astro-ph/0310862)
\re
51.\ Borkowsky K.J., Szymkowiak A.E., Blondin J.M. et al.\ 1996, ApJ, 466,
     866
\re
52.\ Aharonian F.A., Akhperjanian A., Barrio J. et al.\ 2001, A\&A, 370, 112
\re
53.\ Berezhko E.G., P\"uhlhofer G., V\"olk H.J.\ 2003, A\&A, 400, 971
\re

\end{document}